\documentstyle[12pt]{article}
\newcounter{mycount}

\newcommand{\be}{\begin{eqnarray}}
\newcommand{\ee}{\end{eqnarray}}
\newcommand\half{\frac 1 2 }
\newcommand{\sgn}{\mathop{\rm sgn}\nolimits}
\newcommand{\arctanh}{\mathop{\rm arctanh}\nolimits}
\newcommand{\Rh}{\hat{R}}
\newcommand{\qh}{\hat{q}}

\begin{document}

\centerline{\Large\bf Unrealizable Learning in}
\centerline{\Large\bf Binary Feed-Forward Neural Networks}
\vspace* {-35 mm}
\begin{flushright} DRAFT \\ \today \end{flushright}
\vskip 0.9in
\centerline{{\bf M. Sporre$^{1}$}}
\centerline{Nordita}
\centerline{Blegdamsvej 17}
\centerline{DK-2100 Copenhagen \O}
\centerline{Denmark}
\centerline{sporre@nordita.dk}
\vspace{20mm}
\noindent
\centerline{\bf ABSTRACT}
\vspace{5mm}

{\small
Statistical mechanics is used to study unrealizable generalization in two large
feed-forward neural networks with binary weights and output, a perceptron and a
tree committee machine. The student is trained by a teacher being larger, i.e.
having more units than the student. It is shown that this is the same as using
training data corrupted by Gaussian noise. Each machine is considered in the
high temperature limit and in the replica symmetric approximation as well as
for one step of replica symmetry breaking. For the perceptron a phase
transition is found for low noise. However the transition is not to optimal
learning. If the noise is increased the transition disappears. In both cases
$\epsilon _{g}$ will approach optimal performance with a $(\ln\alpha
/\alpha)^k$ decay for large $\alpha$. For the tree committee machine noise in
the input layer is studied, as well as noise in the hidden layer. If there is
no noise in the input layer there is, in the case of one step of repl!
ica symmetry breaking, a phase tra
nsition to optimal learning at some finite $\alpha$ for all levels of noise in
the hidden layer. When noise is added to the input layer the generalization
behavior  is similar to that of the perceptron. For one step of replica
symmetry breaking, in the realizable limit, the values of the spinodal points
found in this paper disagree with previously reported estimates
\cite{seung1},\cite{schwarze1}.
Here the value $\alpha _{sp} = 2.79$ is found for the tree committee machine
and $\alpha _{sp} = 1.67$ for the perceptron.
}

\vspace{5mm}\noindent

\vspace{5mm}

\newpage
\setlength{\baselineskip}{10mm}

\noindent{\Large\bf 1. Introduction }
\renewcommand{\theequation}{1.\arabic{equation}}
\setcounter{equation}{0}

A Feed-forward neural network can be used to estimate an unknown rule from
random examples \cite{hertz1} by adaption of its weights. Using methods from
statistical mechanics of disordered systems \cite{mezard1} the performance of a
student network trained on examples obtained from a teacher network of the same
architecture has been studied (for a review see \cite{watkin1}). In this case
the rule is said to be realizable since it is possible for the student to
develop the same weights as the teacher.

One way to construct an unrealizable rule is to allow for a teacher that is
larger (more units) than the student. This will be shown to be equivalent to
adding Gaussian noise to the training set. The noisy data scenario has been
investigated for networks with continuous weights
\cite{tishby1},\cite{robert1},\cite{robert2}. In the limit where the teacher is
infinitely larger than  the student (large noise limit) the only thing the
student can do is to learn each example by heart, and in this limit the problem
reduces to that of storage capacity.

In this paper the generalization behavior of two different types of binary
neural networks with binary weights is studied, a perceptron (section 2) and a
tree committee machine (section 3), in the limit where the number of units is
large. The rule is defined by a teacher of the same type but having more units
than the student, making the task unrealizable.

The training of the student, having $N$ units, is based on $\alpha N$ examples
obtained by picking inputs $\vec{\xi}^{\mu}$ and assigning outputs $\tau
^{\mu}$ as given by the teacher. With $\sigma ^{\mu}$ being the $\mu$th output
of the student, a training energy
$E = \sum_{\mu} \theta (-\sigma ^{\mu} \tau ^{\mu})$
is defined, which leads to a probability density with Boltzmann weight
$e^{-\beta E}$, where $\beta = 1/T$ is the inverse temperature. First the high
temperature limit is considered for each type of network, the perceptron in
section (2.1) and the tree committee machine in section (3.1). Then, in
sections (2.2) and (3.2), the replica trick is used, assuming replica symmetry
(RS), to study the average over all training sets of the free energy, $\beta
f$. In sections (2.3) and (3.3) the corrections given by one step of replica
symmetry breaking (RSB) are discussed. Since, in the noiseless limit, the value
of the spinodal point found in the RSB-case disagrees with previously reported
estimates \cite{seung1},\cite{schwarze1}, some time is spent on the saddle
point equations in appendix A. Finally, in appendix B, the procedure for
finding the asymptotic generalization behavior for large $\alpha$ is given.

\vspace{1cm}
\noindent{\Large\bf 2. A Large Binary Perceptron with Ising weights.}
\renewcommand{\theequation}{2.\arabic{equation}}
\setcounter{equation}{0}

Let the student and the teacher have $N$ and $M$ input units respectively with
$N\leq M$.
Presented an input, $\vec{s}$, the teacher evaluates,
$\tau (\vec{s}) = \sgn (\vec{v} \cdot \vec{s})$, while the student computes,
$\sigma (\vec{s}_{0}) = \sgn (\vec{w}_{0} \cdot \vec{s}_{0})$,
given the input $\vec{s}_0$. Here $\vec{s}$ and $\vec{v}$ are elements of
${\cal R}^{M}$, while vectors having a zero subscript are elements of ${\cal
R}^{N}$. When the student is presented the same input vector, $\vec{\xi}$, as
the teacher, it only considers the $N$ first components, $\vec{\xi}_{0}$. Thus
the target rule will be,
\be
\tau (\vec{\xi})
&=& \sgn\left[\frac{1}{\sqrt{M}} \sum_{j=1}^{N} v_{j} \xi _{j} +
              \frac{1}{\sqrt{M}} \sum_{j=N+1}^{M} v_{j} \xi _{j}
        \right] \\
&=& \sgn\left[\sqrt{\frac{N}{M}} \left(\vec{v}_{0} \cdot \vec{\xi}_{0} +
	       \frac{1}{\sqrt{N}} \sum_{j=N+1}^{M} v_{j} \xi _{j} \right)
        \right] \\
&=& \sgn\left(\vec{v}_{0} \cdot \vec{\xi}_{0} + \eta\right) \,\,\, ,
\ee
where $\vec{v}_{0}$ is constructed from the first $N$ components of $\vec{v}$.
Effectively this means that the student will be given the task
$\tau' (\vec{\xi}_{0}) = \sgn (\vec{v}_{0} \cdot \vec{\xi}_{0})$
with noise on the input vector
$\vec{\zeta}_{0} = \vec{\xi}_{0} + \vec{\kappa}_{0}$
and/or on the weight vector
$\vec{J}_{0} = \vec{v}_{0} + \vec{\omega}_{0}$.
Since $\eta$ is constructed from independent Gaussian random variables, $v_{j}$
and $\xi _{j}$ ($j=N+1,...,M$) with unit variance, $\eta$ will also be Gaussian
with variance,
\be
\langle \eta ^2 \rangle
&=& \left\langle\frac{M-N}{N}\left(\frac{1}{\sqrt{M-N}}
           \sum_{j=N+1}^{M} v_{j} \xi _{j}\right)^2 \right\rangle\\
&=&\frac{1-\gamma ^2}{\gamma ^2} \,\,\, ,
\ee
where $\gamma = \sqrt{N/M}$. $\gamma$ has the simple interpretation of the
relative size of the student to the teacher. If $\gamma = 1$ the student and
the teacher are of the same size, i.e. there is no noise. If $\gamma=0$ the
teacher is infinitely larger than the student, i.e. the data will be completely
noisy.

The generalization error, $\epsilon _{g}$, obtained by taking the average of
$\theta (-\sigma \tau)$ over normal distributed inputs, $s _{j}$ ($j=1,...,M$),
is
\be\label{eg}
\epsilon _{g} = \frac{1}{\pi} \arccos (\gamma R) \,\,\, ,
\ee
where $R$ is the overlap between $\vec{w}_{0}$ and $\vec{v}_{0}$. For $R=1$ we
obtain the optimal value, $\epsilon _{opt}$, of $\epsilon _{g}$.

First the high temperature limit is considered. Then by using the replica
method, the RS approximation is studied, and finally the corrections given by
one-step RSB are discussed.

\vspace{1cm}
\noindent{\large\bf 2.1. High Temperature Limit}
\renewcommand{\theequation}{2.1.\arabic{equation}}
\setcounter{equation}{0}

In previous work \cite{seung1} the high temperature limit has proven to be
interesting since it is both computationally easy and gives the general
behavior of learning. It is defined so that both $\alpha$ and $T$ approach
infinity while $\alpha \beta$ remains constant. The free energy is simply
\be
\beta f = \frac{\alpha \beta}{\pi} \arccos (\gamma R) +
\frac{1-R}{2} \ln (\frac{1-R}{2}) + \frac{1+R}{2} \ln (\frac{1+R}{2})\,\,\, .
\ee
The qualitative behavior of the learning curves can be divided into two types
depending on whether the noise level is above or below a particular value
$\gamma _0$. For $\gamma _{0} < \gamma < 1$ there is, as in the realizable
case, a range $(\alpha\beta)_{sp1}<\alpha\beta<(\alpha\beta)_{sp2}$, for which
$\beta f$ has two minima. In between $(\alpha\beta)_{sp1}$ and
$(\alpha\beta)_{sp2}$ there is a transition point $(\alpha\beta)_{tr}$ at which
the global properties of the minima change. In contrast to the noiseless case,
$(\alpha\beta)_{sp1} > 0$ and thus for
$0\leq\alpha\beta\leq (\alpha\beta)_{sp1}$ there is only one minimum. The
minimum persisting also for $\alpha\beta\geq(\alpha\beta)_{sp2}$ is close to
$R=1$ and approaches optimal performance as $\alpha\beta$ increase. Note that
in contrast to the realizable case there is no solution at $R=1$. Typically
$(\alpha\beta)_{tr}$, $(\alpha\beta)_{sp1}$ and $(\alpha\beta)_{sp2}$ increase
with decreasing $\gamma$ and merge at $\gamma = \gamma _{0}$. This is
illustrated in figure (\ref{HTLpic}).

The two minima of $\beta f$ must be separated by a maximum, implying that
$\frac{\partial ^2 \beta f}{\partial R^2} =0$
at the spinodal points. Using the saddle point equation,
\be\label{spehtl}
\frac{\alpha\beta\gamma}{\pi \sqrt{1-\gamma ^2 R^2}} =
\half \ln \left(\frac{1+R}{1-R} \right) = \arctanh(R)\,\,\, ,
\ee
to eliminate $\alpha\beta$ from
$\frac{\partial ^2 \beta f}{\partial R^2} =0$,
gives
\be\label{sphtleq}
R=\tanh\left[ \frac{1-\gamma ^2 R^2}{\gamma ^2 R(1-R^2)} \right]
\equiv g(R,\gamma)\,\,\, .
\ee
For $\gamma = 1$, (\ref{sphtleq}) has one solution, $R_{sp}=0.83$ resulting in
$(\alpha\beta)_{sp} = 2.08$ in agreement with \cite{seung1}. In the region
$\gamma _0 <\gamma <1$ (\ref{sphtleq}) has two solutions giving
$(\alpha\beta)_{sp1}$ and $(\alpha\beta)_{sp2}$. At $\gamma = \gamma _0$ the
two solutions merge and the two curves $R$ and $g(R,\gamma)$ are tangent to
each other. Thus $\gamma _0$ can be found by solving,
\be
\frac{\partial}{\partial R} g(R,\gamma _0) &=& 1 \,\,\, ,\\
g(R,\gamma _0) &=& R \,\,\, ,
\ee
giving $\gamma _0 = 0.965$. For $\gamma < \gamma _0$, $\beta f$ has only one
minimum (for all $\alpha\beta$) which moves towards $R=1$ as $\alpha\beta$
approaches infinity.
Note that fairly small amounts of noise will change the qualitative behavior
from phase transition to no transition.

In weight space this behavior can be understood as follows. In the noiseless
case there are, for small $\alpha$, two regions in weight space corresponding
to the minima of $\beta f$, one with poor generalization and one with good. If
$\alpha$ is small enough the ``poor'' region has the lowest free energy. As
$\alpha$ increase the ``poor'' region moves towards the ``good'' and for
$\alpha > \alpha _{tr}$ the ``good'' region has the lowest free energy. Since
for $\alpha = \alpha _{tr}$ the ``poor'' and ``good'' regions are separated,
there will be a phase transition.

If noise is added, the sizes of these regions will increase. For low $\alpha$
there is only one region in weight space corresponding to a minimum of $\beta
f$. It will have
poor generalization. At $\alpha = \alpha _{sp1}$ another region corresponding
to a free energy minimum appears. This region gives better generalization.
Again as $\alpha$ increase the ``poor'' region moves towards the ``good'' and
for
$\alpha > \alpha _{tr}$ the ``good'' region has the lowest free energy. Since
for $\alpha = \alpha _{tr}$ the ``poor'' and ``good'' regions are separated
there  will be a phase transition. If the noise is increased the ``poor''
region is so large that when the ``good'' region is created it will overlap
with the ``poor''. Thus there is only one region, moving towards better
generalization and there is no phase
transition.

\noindent{\large\bf 2.2. Replica Symmetric Theory}
\renewcommand{\theequation}{2.2.\arabic{equation}}
\setcounter{equation}{0}

Using the same methods as in \cite{seung1} the RS approximation to the free
energy is obtained,
\be\label{bfRS}
\beta f_{RS} &=&
\begin{array}{c}{\textstyle extr}\\{\scriptscriptstyle R,\Rh,q,\qh}\end{array}
           \left[G_{r}(R,q,\alpha,\gamma,\beta) +
                 G_{s}(R,\Rh,q,\qh) \right] \,\, ,\\
G_{r} &=&
-2\alpha\int Du \,\, H\left[\frac{\gamma Ru}{\sqrt{q-\gamma ^2 R^2}}\right]
  V\left(u\sqrt{\frac{q}{1-q}}\right) \,\, ,\\
V(x) &=& \ln\left[ e^{\beta}+\left( 1-e^{-\beta}\right)\,\, H(x) \right]\,\,
,\\
G_{s} &=&
\half (1-q)\qh+R\Rh-\int Du\,\,\ln\left[2\cosh\left(\Rh+\sqrt{\qh}
u\right)\right]\,\, .
\ee
The saddle point equations generated by the extremal condition in (\ref{bfRS})
is given in appendix A. Here $q$ is the typical overlap between two different
$\vec{w}_{0}$. $R$, $\gamma$ and $\alpha$ have the interpretation given above.
Using the saddle point equations we can, given $\gamma$ and $\beta$, eliminate
the auxiliary variables $\Rh$ and $\qh$, and find the dependence of $R$ (and
$\epsilon _{g}$) on $\alpha$.

First consider the zero temperature case. This corresponds to only allowing
students that answers all questions correctly. If $\gamma<1$ the training data
is noisy and there is a maximum size of the training set $\alpha _{c}N$ beyond
which no student can perform optimally. $\alpha_{c}(\gamma)$ and
$R_{c}(\gamma)$ are plotted in figure (\ref{critical}).
For $\gamma = 0$ the known result of Gardner \cite{gardner1} is reproduced.
Note that the curves do not give $\alpha_{c} \rightarrow \infty$ as
$\gamma\rightarrow 1$. However this may not be expected since the curves only
give correct predictions for states that are stationary points of $\beta
f_{RS}$ and in the realizable case the state $R=1$ is not stationary as was
shown in \cite{seung1}. For $\gamma = 1$ both the transition and the spinodal
points agrees with the values found in \cite{seung1}. A learning curve for
$\gamma =0.99$ is shown in figure (\ref{RSpic}).

At $T>0$ the learning behavior is the same as for the high temperature limit
but with a different $\gamma _0$, depending on $T$, and with $\epsilon _{g}$
and $q$ having the asymptotic form,
\be\label{asyRS}
\epsilon _{g}-\epsilon _{opt}
&=& C_{1}(\gamma,\beta) \,\, \frac{\ln\alpha}{\alpha}\,\,\, ,\\
1-q
&=& C_{2}(\gamma,\beta) \, \left(\frac{\ln\alpha}{\alpha}\right)^2\,\,\, ,
\ee
for large $\alpha$. For details on how to compute the asymptotic form see
appendix B.
For some range of $\gamma$, $\gamma _{A} <\gamma < 1$, there is a phase
transition already at zero temperature while in a range $\gamma _{B} < \gamma <
\gamma _{A}$ there is no transition at low temperature. As the temperature is
increased a transition develops which is illustrated in figure (\ref{RSpic})
for $\gamma =0.99$.
Finally when
$\gamma < \gamma _{B}$ there seems to be no phase transition no matter how high
the temperature.

\noindent{\large\bf 2.3. Replica Symmetry Breaking}
\renewcommand{\theequation}{2.3.\arabic{equation}}
\setcounter{equation}{0}

In the RS approximation the entropy will always turn negative at some finite
$\alpha$ and therefore a region in $\alpha T$-space for which the system
exhibits replica symmetry breaking (RSB) is expected, see figure (\ref{phase}).
Analogous to \cite{seung1} one step of RSB gives,
\be\label{bfRSB}
\beta f_{RSB} &=&
{\textstyle extr}
       \left[G_{r}(R,q_0,q_1,m,\alpha,\gamma,\beta)+
       G_{s}(R,\Rh,q_0,\qh_0,q_1,\qh_1,m)\right] ,\\
G_{r} &=&
-\frac{2\alpha}{m}\int Dt\int_{0}^{\infty} D\mu \,\,
   \ln\left[
   \int D\omega\left(e^{-\beta}+\left(1-e^{-\beta}\right) H(z)\right)^m
   \right] , \\
z &\equiv&
\frac{t\sqrt{q_0-\gamma ^2 R^2}-\mu\gamma R+\omega\sqrt{q_1-q_0}}{\sqrt{1-q_1}}
\,\, , \\
G_{s} &=&
\half \left( (m-1)q_1 +1 \right) \qh _1 - \frac{m}{2} q_0 \qh _0 + R\Rh
\nonumber \\
      & +&
\frac{1}{m} \int Dt\,\,\ln
\left[\int Dy\left( 2\cosh\left( \Rh+\sqrt{\qh _0}t+
       \sqrt{\qh _1 -\qh_0} y \right) \right)^m
\right] ,
\ee
where the extremum is taken over $R$, $\Rh$, $q_0$, $\qh_0$, $q_1$, $\qh_1$,
and $m$.
As in \cite{seung1} the limit
$q_1\rightarrow 1$, $\qh _1 \rightarrow\infty$ is considered, implying that the
stationary points of $f_{RSB}$ are given by the stationary points of $f_{RS}$
having zero entropy (see appendix A for details).

The learning behavior is analogous to the high temperature limit but with
$\gamma _{0} = 0.995$. In appendix B the asymptotic form of $\epsilon _{g}$ and
$q$ is computed,
\be\label{asyRSB}
\epsilon _{g}-\epsilon _{opt}
&=& C_{3}(\gamma)\, \left(\frac{\ln\alpha}{\alpha}\right)^2 \,\,\, ,\\
1-q
&=& C_{4}(\gamma)\, \left(\frac{\ln\alpha}{\alpha}\right)^2\,\,\, .
\ee
When $0.999\leq \gamma<1$, $\alpha _{tr}$ occurs in between
$\alpha _{sp1}$ and $\alpha _{sp2}$ while for $0.996\leq \gamma\leq 0.998$,
$\alpha _{tr} = \alpha _{sp1}$, i.e. the state with better generalization is
stable as soon as it appears.

For the case $\gamma = 0.05$ the critical capacity, $\alpha _{c}=0.83$
($q_c=0.56$) is found which is compatible with the known results for $\gamma
=0$ \cite{krauth1}.  Some values are given in table (\ref{tabell}) and some
typical learning curves are given in figure (\ref{learn}).

%

In the noiseless limit the result, $\alpha _{sp}=1.67$, correcting a previous
result by Seung {\it et. al.} \cite{seung1} ($\alpha _{sp}=1.63$). The reason
for this is given in appendix A.

It is also interesting to compare with some recently reported upper bounds for
the Ising perceptron \cite{haussler1}. In this article the asymptotical
behavior was found to be the same as (\ref{asyRSB}). The authors found that the
phase transition disappeared below $\gamma = 0.998$ thus not only predicting
the correct qualitative behavior but also giving a tight quantitative bound on
$\gamma _0$. Also, at $\gamma = 0.998$, they found
$\alpha_{tr} = 2.6136$ whereas $\alpha _{tr} = 1.83$ is obtained at $\gamma _0$
given above and using the replica method.

\vspace{1cm}
\noindent{\Large\bf 3. A Binary Committee Machine with Ising weights.}
\renewcommand{\theequation}{3.\arabic{equation}}
\setcounter{equation}{0}

Let the student and the teacher have $N$ ($K$) and $M$ ($L$) input (hidden)
units respectively, with $N\leq M$, and $K\leq L$. We can think of the student
(teacher) as a committee of binary perceptrons each of which has $N/K$ ($M/L$)
input units. As the $l$th perceptron in the teacher is presented an input
$\vec{s}_{l}$ the teacher evaluates,
\be
\tau(\vec s _1,...,\vec s _{L})=
\sgn\left[\frac{1}{\sqrt{L}}\sum_{l=1}^{L}\sgn\left(
\sqrt{\frac{L}{M}}\sum_{m=1}^{M/L} v_{lm} s_{lm}
\right)\right] \,\,\, ,
\ee
while the student computes,
\be
\sigma(\vec s^{\,(0)} _1,...,\vec s^{\,(0)} _{K})=
\sgn\left[\frac{1}{\sqrt{K}}\sum_{k=1}^{K}\sgn\left(
\sqrt{\frac{K}{N}}\sum_{m=1}^{N/K} w^{(0)}_{kn} s^{(0)}_{kn}
\right)\right] \,\,\, ,
\ee
as the $k$th perceptron in the student is given the input $\vec s^{\,(0)} _k$.
Here $\vec s _l$ and $\vec v _l$ are elements of ${\cal R}^{M/L}$ whereas
a zero superscript indicates that the vector is an element of ${\cal R}^{N/K}$.
When the student is presented the same set of input vectors,
$\vec \xi _l$ ($l=1,...,L$), as the teacher it only considers the first $N/K$
components of the first $K$ vectors in that set, $\vec \xi^{\,(0)} _l$
($l=1,...,K$).
Analogous to the simple perceptron we find that this is equivalent to learning
a
noisy target rule,
\be
\tau(\vec \xi^{\,(0)} _1,...,\vec \xi^{\,(0)} _{K})=
\sgn\left[\frac{1}{\sqrt{K}}\sum_{k=1}^{K}\sgn\left(
\sqrt{\frac{K}{N}}\sum_{n=1}^{N/K} v^{(0)}_{lm} \xi^{(0)}_{lm} + \eta _k
\right)+\eta\right] \,\,\, ,
\ee
where $\eta$ and $\eta _k$ are independent Gaussian random variables with
variance,
\be
\langle \eta^2\rangle &=& \frac{L-K}{K} \equiv \frac{1-\gamma ^2}{\gamma ^2}
\,\,\, ,\\
\langle\eta_k^2\rangle &=& \frac{KM}{NL} \equiv\frac{1-\delta ^2}{\delta
^2}\,\,\, .
\ee
$\gamma = \sqrt{K/L}$ is simply the relative number of hidden units of the
student to the teacher while $\delta=\sqrt{NL/(KM)}$ is the relative number of
input units of a perceptron in the student committee to a perceptron in the
teacher committee. Thus $\gamma$ quantifies the noise in the hidden layer and
$\delta$ the noise in the input layer. If $\gamma = \delta =1$ the realizable
case is recovered.

Using these parameters the generalization error is found,
\be
\epsilon _{g} = \frac{1}{\pi} \arccos\left[R_e \right] \,\,\, ,
\ee
where the effective order parameter is given by
$R_e = \frac{2}{\pi}\gamma\arcsin (\delta R)$ and $R$ is the typical overlap
between $\vec w^{(0)}_k$ and $\vec v^{(0)}_k$. Here, analogous to Schwarze and
Hertz \cite{schwarze1}, it is assumed that $R$ is independent of the hidden
unit index $k$.

As for the perceptron case the high temperature limit is considered first. Then
by using the replica method, the RS approximation is studied and finally the
corrections given by one step of RSB are discussed.

\vspace{1cm}
\noindent{\large\bf 3.1. High Temperature Limit.}
\renewcommand{\theequation}{3.1.\arabic{equation}}
\setcounter{equation}{0}

Taking the limits $T\rightarrow\infty$ and $\alpha\rightarrow\infty$ while
keeping $\alpha\beta$ fixed the free energy is found,
\be
\beta f = \frac{\alpha \beta}{\pi} \arccos (R_e) +
\frac{1-R}{2} \ln (\frac{1-R}{2}) + \frac{1+R}{2} \ln (\frac{1+R}{2}).
\ee
If the noise level is low enough, there exists two spinodal points, $\alpha
_{sp1}$ and $\alpha _{sp2}$, with a phase transition in between. In contrast to
the perceptron one find that if there is no input noise ($\delta =1$) there is
a phase transition to optimal performance at a finite $\alpha $ for all values
of $\gamma$. Given a $\gamma$ and that
$\delta _0 (\gamma) < \delta$ a transition to a state approaching optimal
learning in the large $\alpha$ limit is found. For
$\delta < \delta _0 (\gamma)$ the transition vanishes and $\epsilon _{g}$
approaches $\epsilon _{opt}$ as $\alpha$ tends to infinity.
Especially if
$\delta > \delta _{A}=\delta _0(0)$ there is always a phase transition while
for $\delta < \delta _{B}=\delta _0(1)$ there is no phase transition
independent of the hidden noise.
By the same procedure as in section (2.1) one find $\delta _{A} = 0.965$,
$\delta _{B} = 0.924$ and $\delta _0 (\gamma)$ as shown in figure (\ref{dofg}).
Also here $\alpha _{sp1}$, $\alpha _{tr}$ and $\alpha _{sp2}$ increase with
increasing noise.

\vspace{1cm}
\noindent{\large\bf 3.2. Replica Symmetric Theory.}
\renewcommand{\theequation}{3.2.\arabic{equation}}
\setcounter{equation}{0}

Analogous to Schwarze and Hertz \cite{schwarze1} the RS estimate to the free
energy is found,
\be\label{bfRSC}
\beta f_{RS} &=&
\begin{array}{c}{\textstyle extr}\\{\scriptscriptstyle R,\Rh,q,\qh}\end{array}
           \left[G_{r}(R,q,\alpha,\gamma,\beta) +
                 G_{s}(R,\Rh,q,\qh) \right]\,\,\, , \\
G_{r} &=&
-2\alpha\int Du \,\, H\left[\frac{R_e u}{\sqrt{q_e-R_e^2}}\right]
  V\left(u\sqrt{\frac{q_e}{1-q_e}}\right) \,\,\, ,\\
V(x) &=& \ln\left[ e^{\beta}+\left( 1-e^{-\beta}\right)\,\, H(x)\right]\,\,\, ,
\\
G_{s} &=&
\half (1-q)\qh+R\Rh-\int Du\,\,\ln\left[2\cosh\left(\Rh+\sqrt{\qh}
u\right)\right]\,\,\, ,
\ee
where $R_e$ is given above and $q_e = \frac{2}{\pi}\arcsin q$. The value of $q$
is the typical correlation between two $\vec w ^{(0)}_k$ which are assumed to
be independent of the perceptron index $k$. The interpretations of $R$,
$\gamma$, $\delta$ and $\alpha$ are as given above. By using the equations
generated by the extremal condition in (\ref{bfRSC}) to eliminate $\Rh$ and
$\qh$ we can find the dependence of $R$ (and $\epsilon _{g}$) on $\alpha$ given
$\gamma$, $\delta$ and $\beta$.

For $T=0$ one should, as for the perceptron, find a critical capacity, $\alpha
_{c}$, beyond which the student can not perform optimally on the training set.
However, this is not the case implying that the RS-approximation is bad. In the
realizable case the values of both the transition and the spinodal point agree
with \cite{schwarze1}

At $T>0$ the behavior is much the same as in the high temperature limit with
the exception that for $\delta = 1$ the transition is not to an optimal state
but to a state approaching optimal learning as $\alpha$ tends to infinity.
The asymptotical form of
$\epsilon _{g}$ and $q$ for large $\alpha$ can be found for $\delta =1$,
$\gamma < 1$,
\be\label{asyRSC}
\epsilon _{g}-\epsilon _{opt}
&=& B_{1}(\gamma,\beta) \,\, \left(\frac{\ln\alpha}{\alpha}\right)^2 \,\,\, ,\\
1-q
&=& B_{2}(\gamma,\beta) \, \left(\frac{\ln\alpha}{\alpha}\right)^{4} \,\,\, ,
\ee
and for $\delta <1$, $\gamma \leq 1$,
\be\label{asyRSC2}
\epsilon _{g}-\epsilon _{opt}
&=& B_{3}(\gamma,\delta,\beta) \,\,
\left(\frac{\ln\alpha}{\alpha}\right)^{1/3}\,\,\, ,\\
1-q
&=& B_{4}(\gamma,\delta,\beta) \,
\left(\frac{\ln\alpha}{\alpha}\right)^{4/3}\,\,\, .
\ee
The asymptotic behavior can be found by the same method used in appendix B.
As for the perceptron there is  a range of noise levels for which there is no
phase transition at low temperature but one is developed as the temperature is
increased. One such example ($\gamma = 1$, $\delta =0.99$) is used in the phase
diagram (\ref{phaseC}). If the noise is increased above some value there seems
to be no phase transition no matter how high the temperature.

\vspace{1cm}
\noindent{\large\bf 3.3. Replica Symmetric Breaking.}
\renewcommand{\theequation}{3.3.\arabic{equation}}
\setcounter{equation}{0}

As was said in the previous section the RS-approximation fails in predicting a
critical capacity. Also, the entropy will turn negative at some finite $\alpha$
and thus RSB is expected. In figure (\ref{phaseC}) a phase diagram for $\gamma
= 1$, $\delta =0.99$ shows the RSB region.
For one step of RSB, in the limit $q_1 \rightarrow 1$, $\qh _1
\rightarrow\infty$, the free energy is,
\be\label{bfRSBC}
\beta f_{RSB} &=&
\begin{array}{c}{\textstyle extr}\\{\scriptscriptstyle
R,\Rh,q_0,\qh_0,m}\end{array}
           \left[G_{r}(R,q_0,m,\alpha,\gamma,\delta,\beta) +
                 G_{s}(R,\Rh,q_0,\qh _0,m) \right] , \\
G_{r} &=&
-\frac{2\alpha}{m}\int Du \,\, H\left[\frac{R_e u}{\sqrt{q_e-R_e^2}}\right]
  V\left(u\sqrt{\frac{q_e}{1-q_e}}\right)\,\, ,\\
V(x) &=& \ln\left[ e^{\beta m}+\left( 1-e^{-\beta m}\right)\,\,H(x)\right]\,\,
,\\
G_{s} &=&
\frac{m}{2} (1-q_0)\qh_0+R\Rh\nonumber \\
& & -\frac{1}{m}\int Du\,\,\ln\left[2\cosh\left(m\Rh+m\sqrt{\qh}
u\right)\right] \,\, ,
\ee
For reasons analogous to those given in \cite{seung1} for the perceptron the
stationary points of $f_{RSB}$ are given by the stationary points of $f_{RS}$
having zero entropy.

In contrast to the RS-case, but analogous to the high temperature limit, a
transition to optimal learning is found for all $\gamma$ if $\delta =1$.
Using the same notation as in section (3.1) the values of $\delta _{A}$ and
$\delta _{B}$ are $0.9995$ and $0.9847$ respectively, and $\delta _{0}(\gamma)$
is given in figure (\ref{dofgRSB}).

For the case
$\gamma = \delta = 0.05$ the critical capacity, $\alpha _c = 0.95$
($q_c = 0.31$) is found which is compatible with known results for $\gamma =
\delta =0$ \cite{barkai1}.
Typically $\alpha _{sp1}$, $\alpha _{tr}$ and $\alpha _{sp2}$ increase with
increasing unrealizability until $\delta = \delta _{0}(\gamma)$ where
$\alpha _{sp1} = \alpha _{tr} = \alpha _{sp2}$. Some values of
$\alpha _{sp1}$, $\alpha _{tr}$ and $\alpha _{sp2}$ are given in in table
(\ref{tab2}), and some typical learning curves are given in figures
(\ref{learnCd1}) and (\ref{learnCg1}).

As $\alpha\rightarrow\infty$ the asymptotic forms of $\epsilon _{g}$ and $q$
are,
\be\label{asyRSBC}
\epsilon _{g}-\epsilon _{opt}
&=& A_{1}(\gamma,\delta) \,\, \left(\frac{\ln\alpha}{\alpha}\right)^{2/3}\,\,\,
,\\
1-q
&=& A_{2}(\gamma,\delta) \, \left(\frac{\ln\alpha}{\alpha}\right)^{4/3}\,\,\, .
\ee
Appendix B gives details of how to compute the asymptotic behavior, using the
perceptron as an example.
In the realizable limit ($\delta = \gamma =1$) the result $\alpha _{sp} =
2.79$,   correcting the value found in \cite{schwarze1} ($\alpha _{sp} =
2.58$). The reason for the correction is given in appendix A, where the
perceptron is used as an example.

\vspace{1cm}
\noindent{\Large\bf 4. Summary.}
\renewcommand{\theequation}{4.\arabic{equation}}
\setcounter{equation}{0}

In summary we have studied unrealizable learning in two large feed-forward
neural network, a perceptron and a tree committee machine within the replica
symmetric ansatz as well as for one step of replica symmetry breaking. The
average generalization error has been calculated as a function of the load
parameter $\alpha$.

For the perceptron it was shown that using a noisy training set results in a
generalization error approaching optimal learning with increasing $\alpha$
according to a power law of $(\ln \alpha /\alpha)^k$ with $k=2$ in the
RSB-case. If the noise is low enough there is a phase transition at some finite
$\alpha$ to a state which is close to $R=1$. Increasing the noise makes the
transition go away.

For the tree committee machine a similar generalization behavior was found, the
main difference being that there is always a transition to optimal learning at
some finite $\alpha$ if there is no noise in the input layer. Typically, noise
in the input layer gives worse generalization behavior than noise in the hidden
layer. For one step of RSB and with noise in the input layer as well as in the
hidden layer the asymptotic form of $\epsilon _{g}$ was found to be
$(\ln \alpha /\alpha)^k$ with $k=2/3$.

In the realizable cases the values of $\alpha_{sp}$  correct previously
reported results \cite{seung1},\cite{schwarze1}, for the RSB spinodal point in
the two machines. Here $\alpha _{sp} =1.67$ was found for the perceptron and
$\alpha _{sp}=2.79$ for the tree committee machine.

I thank J. Hertz for his valuable advice and direction and R. Urbanczik for
many useful discussions. Also, I would like to thank H. Schwarze for sharing
the code written in connection to ref. \cite{schwarze1} which made it possible
to sort out the disagreement on the spinodal points.

\noindent{\large\bf A. The Saddle Point Equations}
\renewcommand{\theequation}{A.\arabic{equation}}
\setcounter{equation}{0}

In the limit $q_1\rightarrow 1$, $\qh _1\rightarrow\infty$ the one step RSB
free energy (\ref{bfRSB}), of the perceptron, is related to the RS-estimate
(\ref{bfRS}) thru \cite{seung1},
\be\label{frel}
f_{RSB}(R,\Rh,q_0,\qh _0,m,\beta) &=&
    \frac{1}{m} f_{RS}(R,m\Rh,q_0,m^2 \qh _0,\beta m)\,\,\, .
\ee
Stationarity with respect to $R$, $\Rh$, $q_0$ and $\qh _0$ results in the
relations $q_0(T_{RSB},m,\alpha) = q_{RS}(T_{RSB}/m,\alpha)$ and
$R(T_{RSB},m,\alpha) = R_{RS}(T_{RSB}/m,\alpha)$ while stationarity with
respect to $m$ gives $s_{RS}(T_{RSB}/m,\alpha)=0$ where $s_{RS}$ is the RS
entropy. Thus one can find the stationary points of $f_{RSB}$ by finding
stationary points of $f_{RS}$ at a temperature $T_{RS}=T_{RSB}/m$ for which the
entropy is zero. The saddle point equations generated by (\ref{bfRS}) are
\be
q &=& \int Dt\,\, \tanh ^2(\Rh+\sqrt{\qh}t) \label{speq}\,\,\, ,\\
R &=& \int Dt\,\, \tanh (\Rh+\sqrt{\qh}t) \label{speR}\,\,\, ,\\
\qh &=&
\frac{\alpha}{\pi(1-q)}
\int Du \,\,H\left[\frac{\gamma R u}{\sqrt{q-\gamma ^2 R^2}}\right]
\frac{e^{-v^2}}{\left[u_{\beta} + H(v)\right]^2} \label{speqh}\,\,\, ,\\
\Rh &=&
\frac{\alpha}{\pi\sqrt{1-q}}
\int Dt\,\, \frac{e^{-y^2/2}}
             {u_{\beta}+H(y)}\label{speRh}\,\,\, ,
\ee
with $u_{\beta}=1/(e^{\beta}-1)$, $v=u\sqrt{q/(1-q)}$ and
$y=t\sqrt{(q-\gamma ^2 R^2)/(1-q)}$.
Using (\ref{speq}) and (\ref{speR}) to eliminate $q$ and $R$ in (\ref{speqh})
an (\ref{speRh}) gives a system of two non-linear equations
\be
\qh &=& \alpha \,\, h(\Rh,\qh) \label{twoqh}\,\,\, ,\\
\Rh &=& \alpha \,\, g(\Rh,\qh) \label{twoRh}\,\,\, .
\ee
At this point we could try to solve for $\qh$ and $\Rh$ given $\alpha$. However
since $\epsilon _{g}$ is a many-valued function of $\alpha$  it is more
economical to eliminate $\alpha$. This will give the equation,
\be\label{oneq}
\qh\,\, g(\Rh,\qh) = \Rh \,\, h(\Rh,\qh)\,\,\, ,
\ee
which can be solved for $\qh$ given $\Rh$. $\alpha$ can be evaluated using
(\ref{twoqh}) or (\ref{twoRh}). The advantage is that $\epsilon _{g}$ is a
single valued function of $\Rh$. In the RSB-case this will be helpful since
more than one solution for each $\alpha$ has to be considered as we show below.

Once a stationary point has been found its second order properties has to be
checked by computing the determinant of the Hessian matrix, $H$. Assume that
the correct sign of $\det H$, at $\Rh >0$, is given by the sign at $\Rh =0 $.
As $\Rh$ is increased, the sign of $\det H$ will change first at $\Rh _{sp2}$
and then again at $\Rh _{sp1}$. Note that $\Rh _{sp2} < \Rh _{sp1}$ whereas
$\alpha _{sp1} < \alpha _{sp2}$. In the regime
$\Rh _{sp2}<\Rh <\Rh _{sp1}$, $\beta f$ has a stationary point but it has the
incorrect curvature.

Even though the RSB-case is solved by using the RS-equations the determinant of
the Hessian matrix of $\beta f_{RSB}$, $\det H_{RSB}$, has to be used to
determine the second order properties. $\det H_{RSB}$ consists of the second
derivatives of $\beta f_{RSB}$ with respect to $R$, $q$, $\Rh$, $\qh$ and $m$
whereas $\det H_{RS}$ is computed from the second derivatives of $\beta f_{RS}$
with respect to $R$, $q$, $\Rh$ and $\qh$. Using $\det H_{RS}=0$ as the
criterion to determine the spinodal point (at $\gamma =1$) would result in the
values of $\alpha _{sp}$ as given in \cite{seung1} and \cite{schwarze1}.
Moreover, insisting on this RS-criterion will, for some $\gamma$, result in
regions of $\alpha$ where no solution exist. Thus this procedure fails in a
disastrous way.
However the correct condition, $\det H_{RSB}=0$, will cure this problem and
give
$\alpha _{sp}$ as given in this paper.

In the RSB-case $\det H_{RS}$ will have the wrong sign close to $\alpha _{sp}$.
This will correspond to points on $\epsilon _{g}$ not considered in the RS-case
since there exists another solution (with $s>0$) at the same $\alpha$ but with
$\det H_{RS}$ having the correct sign. This is illustrated (for $\gamma =1$) in
figure (\ref{rsbspin}).

When the RS-equations are used to solve the RSB-case it is not possible to find
the value of $m$ (only $T_{RS}=T_{RSB}/m$). Since $\det H_{RSB}$ depends on $m$
it can not be computed. However it is possible to show that,
\be
\det H_{RSB}(R,\Rh,q_0,\qh _0,m) = \frac{1}{m} F(R,\Rh,q_0,\qh _0) \,\,\, ,
\ee
and since $0<m<1$, $\det H_{RSB}$ has the same sign as $F$.

\noindent{\large\bf B. Asymptotics}
\renewcommand{\theequation}{B.\arabic{equation}}
\setcounter{equation}{0}

For large $\alpha$ the saddle point equations (\ref{speq}) implies that $q,R$
are close to 1 and $\qh,\Rh$ are large. From this the asymptotic form of the
free energy (\ref{bfRS}) for non-zero temperatures can be found,
\be\label{bfRSasy}
\beta f_{RS} &=& \begin{array}{c}{\textstyle extr}\\{\scriptscriptstyle
R,\Rh,q,\qh}\end{array}
           \left[G_{r}(R,q,\alpha,\gamma,\beta) +
                 G_{s}(R,\Rh,q,\qh) \right] \,\, ,\\
G_{s} &=& \frac{1}{2}(1-q)\qh +R\Rh -\sqrt{\frac{2}{\pi}}\sqrt{\qh}
\exp\left(-\frac{\Rh ^2}{2\qh}\right) - \Rh\left[1-2
H\left(\frac{\Rh}{\sqrt{\qh}}\right)\right] \,\, , \\
G_{r} &=& \frac{\alpha\beta}{\pi}\arctan\left[\frac{\sqrt{q-\gamma ^2
R^2}}{\gamma R}\right] - \frac{\alpha[\phi (\beta) + \phi
(-\beta)]}{\sqrt{2\pi}}\sqrt{1-q} \,\, , \\
H(x) &=& \int_{x}^{\infty} \frac{dy}{\sqrt{2\pi}} e^{-\frac{1}{2}y^2} \,\, , \\
\phi(\beta) &=& \int_{0}^{\infty} dw \ln\left[1+(e^{\beta}-1)H(w)\right] \,\, .
\ee
If $\gamma <1$ (\ref{bfRSasy}) generates the saddle point equations,
\be
1-R &=& \sqrt{\frac{2}{\pi}}\sqrt{\frac{\qh}{\Rh^2}}\exp\left(-\frac{\Rh
^2}{2\qh}\right) \label{asyspeqR}\,\, ,\\
1-q &=& \sqrt{\frac{2}{\pi}}\frac{1}{\sqrt{\qh}}\exp\left(-\frac{\Rh
^2}{2\qh}\right) \label{asyspeqq}\,\, ,\\
\qh &=& \frac{\alpha\beta\gamma}{\pi\sqrt{1-\gamma^2}} +
        \frac{\alpha[\phi (\beta) + \phi (-\beta)]}{\sqrt{2\pi}\sqrt{1-q}}
\label{asyspeqqh} \,\, , \\
\Rh &=& \frac{\alpha\beta\gamma}{\pi\sqrt{1-\gamma^2}}\label{asyspeqRh} \,\, .
\ee
The first two of these can be combined into,
\be
\frac{1-R}{1-q} = \frac{\qh}{\Rh} \label{sepcspeq} \,\, .
\ee
Using (\ref{asyspeqR}) and (\ref{asyspeqqh})-(\ref{sepcspeq}) results in,
\be
1-q &\sim & (1-R)^2 \label{onemqeq}\,\, ,\\
(1-R)^{3/2} &\sim& \frac{1}{\sqrt{\alpha}}\exp\left(-\alpha A_2 (1-R)\right)
\label{onemReq} \,\, ,
\ee
where $A_2$ depend only on $\beta$ and $\gamma$ and where $\sim$ means
proportional to in the asymptotic limit of large $\alpha$. In order to solve
(\ref{onemReq}) the ansatz,
\be\label{ansatzR}
1-R(\alpha) = A_0 \frac{\ln \alpha}{\alpha} + \delta (\alpha) \,\, ,
\ee
is made.
For consistency it is important to check that $\delta (\alpha)$ is of lower
order than $\ln (\alpha)/\alpha$. Combining (\ref{onemReq}) with the ansatz
(\ref{ansatzR}) and by choosing $A_0 = 1/A_2$ gives the solution,
\be\label{asysol}
\epsilon _{g} - \epsilon _{opt} \sim 1-R \sim \frac{\ln \alpha}{\alpha} \,\, ,
\ee
Also $\delta (\alpha) \sim \ln [ln(A_0 \ln\alpha)/A_2^{2/3}]/\alpha$ is found
and thus $\delta (\alpha)$ is of lower order. The asymptotic form of $1-q$ is
now easily found using (\ref{asysol}) and (\ref{onemqeq}).

In the RSB-case the temperature is given by the zero entropy condition and can
not be regarded as an arbitrary constant. Thus $\beta$ is a function of
$\alpha$ and combining the saddle point equations
(\ref{asyspeqR})-(\ref{sepcspeq}) with the asymptotic form of the zero entropy
condition, $\qh \sim (1-q)^{-1/2}$, gives,
\be
1-q &\sim & \beta ^2 \label{qasy}\,\, ,\\
1-R &\sim & \beta ^2 \label{Rasy}\,\, , \\
\beta ^{5/2} &\sim& \frac{1}{\sqrt{\alpha}} \exp\left(-B_2 \alpha \beta\right)
\,\, ,\\
\ee
where $B_2$ only depend on $\gamma$. Again an ansatz,
$\beta (\alpha) = B_0\ln (\alpha)/\alpha +\delta (\alpha)$ is made which
together with $B_0 = 2/B_2$ gives the solution,
\be
\beta (\alpha ) \sim \frac{\ln \alpha}{\alpha} \,\, .
\ee
The asymptotic form of $R$, $q$ and $\epsilon _{g}$ is now found from
(\ref{qasy}) and (\ref{Rasy}) giving,
\be
\epsilon _{g} - \epsilon _{opt} \sim 1-R \sim \left(\frac{\ln
\alpha}{\alpha}\right)^2 \,\, ,
\ee

For the tree committee machine the asymptotic forms of $R$, $q$ and $\epsilon
_{g}$ can be found by the same procedure but using the asymptotic form of the
free energy (\ref{bfRSC}).

\end{document}